\documentclass[letterpaper,prl,twocolumn,preprintnumbers]{revtex4}
\usepackage{graphicx}
\begin{document}

\newcommand{\mysw}[1]{{\scriptscriptstyle #1}}
\newcommand{\mybm}[1]{\mbox{\boldmath$#1$}}
\renewcommand{\thefootnote}{\alph{footnote}}
\title{Spiral Exchange Interaction in Diluted Magnetic Semiconductor Junction}
\author{Shih-Jye Sun}
\affiliation{Department of Applied Physics, National Chia-Yi
University, Chia-Yi 600, Taiwan}

\author{Song-Shien Cheng}
\affiliation{Department of Physics, National Tsing-Hua University,
Hsinchu 300, Taiwan}

\author{Hsiu-Hau Lin}
\affiliation{Department of Physics, National Tsing-Hua University
and Physics Division, National Center for Theoretical Science,
Hsinchu 300, Taiwan}
\date{\today}

\begin{abstract}
We studied the exchange interactions mediated by polarized
itinerant carriers in diluted magnetic semiconductor (DMS)
junction. In contrast to the ordinary RKKY oscillations, the
induced moment possesses an interesting spiral motion, accompanied
with angular oscillations. The spiral motion remains robust in the
entire $T<T_{c}$ regime while the oscillatory motion gets enhanced
as $T \to T_{c}$. To explore the novel phenomena, we propose a
ferromagnet/DMS/ferromagnet junction would bring out interesting
spin-dependent transport properties and possible applications in
spintronics.
\end{abstract}
\maketitle

One of the key elements in spintronics is the diluted magnetic
semiconductor (DMS), which incorporate both charge and spin
phenomena together in one material\cite{Ohno98a,Prinz98}. With
rapid and intensive research attentions, robust ferromagnetic
order was demonstrated in (Ga$_{1-x}$Mn$_x$)As with Currie
temperature around 150 K\cite{Ku03,Edmonds02}. It is generally
believed that the ferromagnetism in DMS is mediated by the
itinerant carriers in the host
semiconductor\cite{Akai98,Konig00,Schlienmann01,Litvinov01}. For
the unpolarized itinerant carriers, the mediated exchange coupling
among magnetic moments is the well-known RKKY interaction.
However, since itinerant carriers in DMS is polarized with
sensitive temperature dependence\cite{Ohno98b,Matsumoto,Dietl1},
it is interesting to study what kind of exchange coupling gets
generated.

In this Letter, we employ the self-consistent Green's function approach to
address this issue. Note that, while the RKKY exchange interaction is known
for a long time, its direct experimental evidence is only detected recently\cite{Hindmarch03}.
An easier way is through the spin-dependent scattering in the ferromagnet/nonmagnetic metallic
junction\cite{Baibich88,Binasch89}. Here we proposed a similar experimental setup, as shown in Fig. 1,
 to explore the effective exchange interaction in DMS. One might expect the effective exchange
 interaction mediated by itinerant carriers in DMS also exhibits similar RKKY-type behavior.
 However, we found the presence of finite polarization gives rise to qualitatively different behavior.
 In the ferromagnet/DMS/ferromagnet junction, the induced polarization is spiral with angle $\theta_{s}(d)$
 depending on the width of the junction. The spatial dependence of the spiral angle at different temperatures
 will be studied in details later.

\begin{figure}
\centering
\includegraphics[width=6cm]{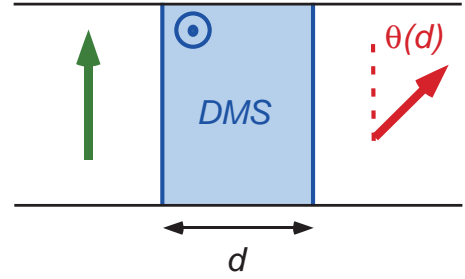}
\caption{Schematic diagram for a
ferromagnet/DMS/ferromagnet junction. The polarization of DMS is
placed in the $z$-direction and the magnetization on the left is
fixed in $y$-direction. The induced polarization on the right has
a spiral angle $\theta(d)$ depending on the junction width $d$.}
\end{figure}

The DMS is modeled by the Hamiltonian, containing the kinetic energy of the itinerant carriers and the
exchange interaction between itinerant carriers and the localized moments doped into the host semiconductor,
\begin{equation}
H= H_{0} + J \int d^{3}r \mybm{S}(r) \cdot \mybm{s}(r),
\label{kondo}
\end{equation}
where the spin density of the localized moments is $\mybm{S}(r) \sum_{I} \delta^{3}(r-R_{I}) \mybm{S}_{I}$
and the itinerant spin density is $\mybm{s}(r) = \psi^{\dag}(r) (\mybm{\sigma}/2) \psi(r)$.
The band structure of the itinerant carriers is described by the host semiconductor.
Since our emphasis here is to demonstrate the qualitative difference from the usual
RKKY exchange coupling, the dispersion is taken to be the simplest parabolic band $H_{0} = p^2/2m^*$.
Generalization to more realistic but complex band structures, such as the six-band Luttinger model,
can be achieved straightforwardly. Throughout the Letter, the parameters would be set to the typical
values in (Ga$_{1-x}$, Mn$_{x}$)As: $J = 0.15$ eV nm$^3$ and $m^*=0.5 m_{e}$. Besides,
the densities of itinerant carriers and localized moments are $c^*=0.1$ nm$^{-3}$ and $c = 1$ nm$^{-3}$ respectively.

The induced exchange interaction $J_{ij}(r)$ mediated by polarized itinerant carriers is directly
proportional to the static spin susceptibility, $J_{ij}(r) = J^2 \chi_{ij}(r, \omega=0)$,within the
linear-response regime. The spin susceptibility is
\begin{eqnarray}
\chi_{ij}(r,t) = -i \Theta(t) \langle\langle [s_{i}(r,t), s_{j}(0,0)] \rangle\rangle,
\label{Susceptibility}
\end{eqnarray}
where the double brackets implies both thermal and quantum averages. To compute the susceptibility
at finite temperature, we employed the self-consistent Green's function method developed
previously\cite{Sun03,Sun02} to obtain the propagators for itinerant carriers and localized moments.
With these Green's functions, the average spin densities of itinerant and localized spins,
$\langle S_{z} \rangle$ and $\langle s_{z} \rangle$ can be determined self-consistently in numerics.
With the particular set of parameters we choose in this Letter, the Curie temperature is about $T_{c} \approx 45$ K.

Making use of Eq.~\ref{Susceptibility} from employing equation of
method on it\cite{doniach}, the exchange couplings generated from
the diagonal parts of the spin susceptibility are
\begin{eqnarray}
\label{jz}
J_{zz}(r)&=&\frac{J^{2}m}{4\pi^{3}\hbar^2}\frac{1}{r^2}
\int^{\infty}_{0} k\sin(kr)\cos(kr)(f_{\uparrow}+f_{\downarrow})dk,
\\
J^{xx}(r)&=&\frac{J^{2}m}{4\pi^{3}\hbar^2} \frac{1}{r^2} \bigg[
\int^{\infty}_{0} k\sin(kr)\cos(\sqrt{k^2+k_{\Delta}^2} r) f_{\downarrow} dk
\nonumber\\
&+& \int^{\infty}_{k_{\Delta}} k\sin(kr)\cos(\sqrt{k^2-k_{\Delta}^2} r) f_{\uparrow} dk
\nonumber\\
&+& \int^{k_{\Delta}}_{0} k\sin(kr) e^{-\sqrt{k_{\Delta}^2 -k^2}r} f_{\uparrow} dk \bigg],
\end{eqnarray}
where $\hbar^2 k^2_{\Delta}/2m^* =\Delta$ and $\Delta = J \langle S_{z} \rangle$ is the effective
Zeeman gap caused by the average local magnetization. The Fermi-Dirac distribution for spin-up and
spin-down itinerant carriers are $f_{\uparrow,\downarrow}$ respectively.

\begin{figure}
\centering
\includegraphics[width=6cm]{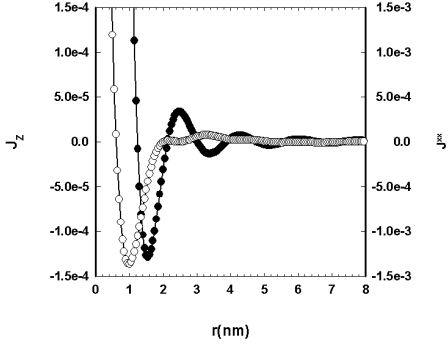}
\caption{The diagonal parts of exchange interactions
$J_{zz}(r)$ (filled symbols) and $J_{xx}(r)$ at $T=10K$.}
\end{figure}

The spatial dependences of $J_{zz}(r)$ and $J_{xx}(r)=J_{yy}(r)$ are shown in Fig. 2. It is evident
that they show oscillatory RKKY behavior and decay quickly from the origin. Note that the finite
polarization causes the non-uniform decay of the oscillation amplitudes. The finite polarization also
gives rise to the off-diagonal exchange interactions. It is straightforward to show that $J_{zx}=J_{zy}=0$
from symmetry arguments. The only non-vanishing off-diagonal exchange interaction is
\begin{eqnarray}
J^{xy}(r)&=&\frac{J^{2}m}{4\pi^{3}\hbar^2} \frac{1}{r^2}
\int_{k_{\Delta}}^{\infty} k \sin(kr) \sin(\sqrt{k^2-k^2_{\Delta}}r) f_{\uparrow} dk
\nonumber\\
&-& \int_{0}^{\infty} k \sin(kr) \sin(\sqrt{k^2+k^2_{\Delta}}r) f_{\downarrow} dk.
\end{eqnarray}
It is clear that the first and second integrals cancel when the polarization vanishes $k_{\Delta} =0$
and $f_{\uparrow} = f_{\downarrow}$. The spatial dependence of $J_{xy}(r)$ is also oscillatory as the diagonal parts.
However, the existence of the off-diagonal exchange interaction gives rise an interesting phenomena
which is absent for usual RKKY interaction.

\begin{figure}
\centering
\includegraphics[width=7cm]{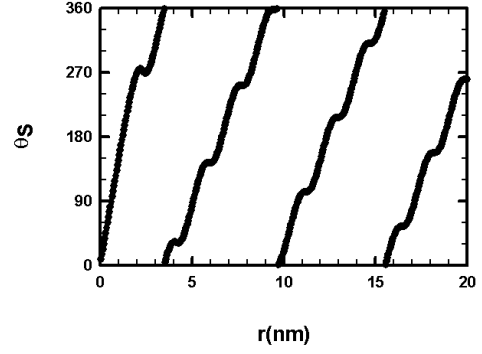}
\caption{The spiral angle $\theta_{s}(r)$ at $T=10$ K.
The chemical potential only cut across the single band of one
particular spin flavor and the polarization of itinerant carriers
is almost one.}
\end{figure}

Suppose we sandwich DMS with two ferromagnets as shown in Fig. 1.
The perpendicular polarization between DMS and the left
ferromagnet could be made by utilizing the anisotropy of DMS
making the hard axis of DMS parallel to the left ferromagnet. The
moment on the left ferromagnet would induce magnetization in both
$y$- and $x$-axes since the off-diagonal part is not zero. It is
useful to define the spiral angle of the induced magnetic moment,
\begin{eqnarray}
\tan \theta_{s}(r) = \frac{J_{xy}(r)}{J_{yy}(r)}.
\end{eqnarray}
The spatial dependences of the spiral angle at $T=10$ K and $T=44$ K are plotted in Figure 3 and 4.
It is rather interesting that the induced moments through the DMS junction is characterized by both spiral
rotations and angular oscillations. After some algebra, the corresponding length scales of spiral and
oscillatory motions are approximately
\begin{eqnarray}
\lambda_{s} &\approx& \frac{2\pi}{\sqrt{k^2_{F}+k^2_{\Delta}} - k_{F}},
\label{Spiral}
\\
\lambda_{o} &\approx& \frac{2\pi}{\sqrt{k^2_{F}+k^2_{\Delta}} + k_{F}},
\label{Oscillatory}
\end{eqnarray}
where $k_{F}$ is the Fermi momentum of the unpolarized itinerant carriers.

At low temperature $T=10$ K, the chemical potential only cuts
through single band and most of the itinerant carriers have the
same spin orientation. The direction of the induced moment points
along the direction of the spiral angle $\theta_{s}$, depending on
the width of the DMS junction. As is clear from Fig. 3,
accompanied with the spiral motion, there is also the oscillatory
motion (with shorter period) and creates the wiggles in the trace.
A remarkable feature is that the spiral motion does not seems to
decay away as quickly as the exchange couplings $J_{xx}(r)$ and
$J_{xy}(r)$ themselves. This may imply that the spiral dependence
is easier to observe in experiments than the usual RKKY
oscillations. Besides, the spiral-angular dependence of the
induced moment also makes it charming for spin-dependent transport
in tunnelling magnetoresistance (TMR) application and open up a
new window for new functionality of device designs.

Close to the Currie temperature $T=44$ K, both bands of the itinerant carriers are almost degenerate
and the polarization is small. From Eq.~\ref{Spiral}, we expect that the period of spiral motion become
large, as shown in Fig. 4. The oscillatory motion becomes transparent within one spiral period and decays
away rapidly. While one might naively expect that the spiral motion is weaken by the small polarization,
it is actually not the case. The spiral motion persists but with a longer period. As the temperature
approaches the critical one, the period of the spiral motion becomes infinitely long
while the oscillatory motion gets enhanced. Eventually, the spiral angle only takes on discrete values,
either $\theta_{s}=0$ or $\theta_{s}=\pi$, which correspond to the ordinary RKKY oscillations. However,
as long as $T<T_{c}$, the mediated exchange interaction is qualitatively different from that generated
by unpolarized itinerant carriers.

\begin{figure}
\centering
\includegraphics[width=7cm]{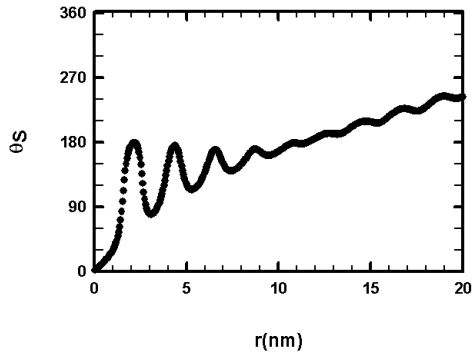}
\caption{The spiral angle $\theta_{s}(r)$ at $T=44$ K.
Both bands of different spins are almost degenerate and the spin
polarization is small.
}
\end{figure}

In conclusion, we studied the exchange interactions mediated by
polarized itinerant carriers in DMS junction. We found, in
contrast to the ordinary RKKY oscillations, the induced moment
possesses a spiral motion accompanied with oscillatory motions.
The spiral motion remains robust in the entire regime below the
Curie temperature, while the angular oscillations get enhanced as
the polarization becomes small. A ferromagnet/DMS/ferromagnet
junction would bring out this novel spiral dependence and leads to
interesting spin-dependent transport phenomena.

We thank the support of National Science Council in Taiwan through
grants NSC-91-2112-M-230-001(SJS), NSC-90-2112-M-007-050(HHL),
NSC-91-2112-M-007-040(HHL) and NSC-91-2120-M-007-001(HHL). The
hospitality of the National Center for Theoretical Science, where
the work was initiated, is greatly acknowledged.

\end{document}